\documentclass[aps, 10pt, pra, twocolumn, groupaddress]{revtex4-1}

\usepackage{amsmath,xcolor}
\usepackage{amsfonts}
\usepackage{amssymb}
\usepackage{graphicx}
\usepackage{srcltx}
\usepackage{mathcomp}   
\usepackage{ulem}       
\usepackage[mathscr]{euscript}
\usepackage{mathrsfs}
\usepackage{tabularx,ragged2e,booktabs}

\newcommand{\beq}{\begin{equation}}
\newcommand{\eeq}{\end{equation}}
\newcommand{\bea}{\begin{eqnarray}}
\newcommand{\eea}{\end{eqnarray}}
\newcommand{\ep}{\epsilon}

\begin{document}

\title{Experimental measurement-device-independent quantum digital signatures}

\author{G.~L.~Roberts, M.~Lucamarini, Z.~L.~Yuan, J.~F.~Dynes, L.~C.~Comandar, A.~W.~Sharpe, A.~J.~Shields}
\affiliation{Toshiba Research Europe Ltd, 208 Cambridge Science Park, Cambridge CB4 0GZ, United Kingdom}
\author{M. Curty}
\affiliation{EI Telecomunicaci\'on, Department of Signal Theory and Communications, University of Vigo, Vigo E-36310, Spain}
\author{I.~V.~Puthoor, E.~Andersson}
\affiliation{SUPA, Institute of Photonics and Quantum Sciences, Heriot-Watt University, Edinburgh EH14 4AS, United Kingdom}

\begin{abstract}
\noindent We propose and experimentally implement a novel reconfigurable quantum key distribution (QKD) scheme, where the users can switch in real time between conventional QKD and the recently-introduced measurement-device-independent (MDI) QKD. Through this setup, we demonstrate the distribution of quantum keys between three remote parties connected by only two quantum channels, a previously unattempted task. Moreover, as a prominent application, we extract the first quantum digital signature (QDS) rates from a network that uses a measurement-device-independent link. In so doing, we introduce an efficient protocol to distil multiple signatures from the same block of data, thus reducing the statistical fluctuations in the sample and increasing the final QDS rate.
\end{abstract}

\maketitle

\section{Introduction}
\label{sec:intro}

\noindent {\Large C}ryptography is ubiquitous in modern society and essential to countless applications relying on the confidentiality, integrity and non-repudiation of sensible data~\cite{MVV05}. Currently, the security of these applications is largely based on public-key cryptography~\cite{DH76,RSA78}, which is supposedly secure against an eavesdropper with limited computational power. QKD, on the other hand, poses no restrictions on the attacker, apart from obeying the laws of Nature, and one only makes assumptions on the devices owned by the authorised users, which can be directly tested. In addition, the recent introduction of MDI-QKD~\cite{BP12,LCQ12} further enhances the positive features of QKD. By clever use of the teleportation gate~\cite{BBC+93}, MDI-QKD turns the receiving side of QKD into a transmitter, thus removing all the security assumptions on the detecting devices, which are arguably most exposed to external attacks~\cite{GLL+11,LWW+10,XQL10,ZFQ+08,QFL+07}. Moreover, it allows two parties to connect through a totally untrusted node. Recent experiments have shown that it can be implemented with key rates commensurate to those of QKD~\cite{CLF+16}, whilst extending its transmission distance~\cite{YCY+16}.

Different from encryption, quantum digital signatures (QDS) have also recently been introduced~\cite{GC01,ACJ06,qds}. Quantum signatures allow users to sign a document by quantum means, and to transfer it to other users, with information-theoretical security.
The most recently proposed quantum signature schemes~\cite{AWK+16} only need experimental setups similar to those for QKD. In analogy with MDI-QKD, measurement-device-independent quantum signatures have also been proposed~\cite{PAW+16}.
\begin{figure}
  \centering
  \includegraphics[width=\columnwidth]{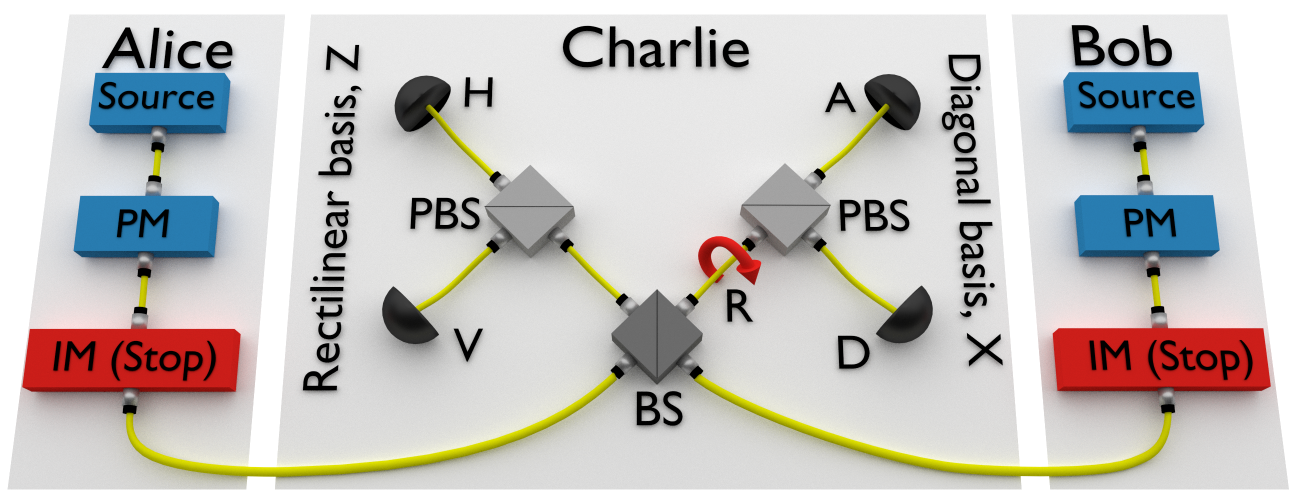}\\
  \caption{\textbf{Fiber-based reconfigurable MDI/QKD network}. The key elements are coloured red. The rotator (R) in Charlie's station sets each detector to measure a different polarization state, $H$, horizontal, $V$, vertical, $D$, diagonal, and $A$, anti-diagonal. The intensity modulators (IMs) can be set to high attenuation, to nearly stop the light passing through them (``Stop'' in the figure). This can enable Alice-Bob MDI-QKD, when no IM is set to Stop, or Alice-Charlie QKD, when Bob's IM is set to Stop, or Bob-Charlie QKD, when Alice's IM is set to Stop. At the same time, the IMs can be used to prepare decoy states~\cite{dec1,dec2,dec3,dec4}. PM: polarization modulation; BS: beam splitter; PBS: polarising beam splitter.}
  \label{fig1}
\end{figure}

Therefore, various quantum communication schemes are already available and the specific application determines which is the most suitable in a given situation. In such a scenario, it becomes important to be able to switch between different protocols, in order to adapt to a specific demand. For instance, let us consider the basic three-point network depicted in Fig.~\ref{fig1}, where two distant users, Alice and Bob, are connected through a central node, Charlie. In some cases, Alice and Bob might want to communicate privately through Charlie's relay station. For that, they can run the decoy-state MDI-QKD protocol~\cite{LCQ12} using the equipment shown in Fig.~\ref{fig1}. In particular, the two intensity modulators (IM) allow them to implement the decoy-state technique~\cite{dec1,dec2,dec3,dec4}. The resulting key will be unknown to Charlie and to any external eavesdropper, and secure against attacks directed at Charlie's equipment. In other cases, Alice and Bob might want to communicate directly with Charlie, who then represents a legitimate user and can exchange keys with them via QKD. In the scheme in Fig.~\ref{fig1}, a QKD transmission between Alice (Bob) and Charlie is obtained by stopping the light emitted by Bob (Alice) through the same IMs employed for the decoy-state MDI-QKD protocol. If the users' devices are trusted, the resulting QKD transmission will feature quantum security against external eavesdroppers and it will be fast, with megabits of key material distributed every second. The possibility to switch between QKD and MDI-QKD constitutes a ``reconfigurable MDI/QKD network'', a concept introduced in~\cite{QLL+15} for free-space quantum communications.

In this work, we experimentally realise the fiber-based reconfigurable MDI/QKD network sketched in Fig.~\ref{fig1}. By sending light to the central beam-splitter (BS), Alice and Bob can distil a key via MDI-QKD, whereas by stopping one of the light beams, Alice or Bob can distil keys with Charlie via QKD. Stopping the light beams is simple in our scheme, as it is effected through the same intensity modulators employed to implement the BB84 protocol~\cite{BB84} with decoy states~\cite{dec1,dec2,dec3}, both in QKD~\cite{dec4} and in MDI-QKD~\cite{LCQ12}. The obtained key rates range from hundreds of bits per second (bps) to hundreds of kilobits per second (kbps) for MDI-QKD, and from hundreds of kbps to millions of bps (Mbps) for QKD.

In addition, the setup can be used in a way different from standard QKD and MDI-QKD, to realise quantum signatures.
As a striking example of this, we experimentally estimate the first QDS rates mediated by an QKD/MDI-QKD setup. The resulting rates are in the order of 1 signature in 45 seconds for the MDI-QKD link and 1 signature in 72~ms for a QKD link used for signature distribution, the former referring to a 50 km optical fiber, the latter to a 25 km one. QKD can be a point-to-point application with only two users in the simplest scenario. For quantum signatures, on the other hand, three parties, all pairwise connected, is the minimal scenario. Our QKD/MDI-QKD setup implements the three required quantum communication links using only two physical links.

\section*{Experimental setup}
\label{sec:expsetup}

\noindent To implement the fiber-based reconfigurable MDI/QKD network, we adopt a polarization-based setup. We denote by $H$, $V$, $D$ and $A$ the horizontal, vertical, diagonal and anti-diagonal states of linear polarization, respectively, and with $Z$ and $X$ the rectilinear and diagonal bases, composed of the states $\{H,V\}$ and $\{D,A\}$, respectively. The setup makes use of the decoy-state technique to improve the key rate and extend the transmission distance. Therefore the preparation step also includes the selection of the intensity of the pulses to be sent to Charlie. In this case, we adopt the scheme with four intensity classes~\cite{ZYW16,CLF+16}, indicated as $s$ (``signal''), $u$ (``decoy1''), $v$ (``decoy2'') and $w$ (``vacuum''). The signal $s$ is the only one prepared in the $Z$ basis, whereas $u$, $v$ and $w$ are all prepared in the $X$ basis. The quantum keys and signatures are extracted from the $s$ pulses in the $Z$ basis, whereas the $X$ basis is for testing the quantum channel against the presence of an eavesdropper. To increase the final key rate, the basis $Z$ is selected more often than $X$.

\begin{figure}
  \centering
  \includegraphics[width=\columnwidth]{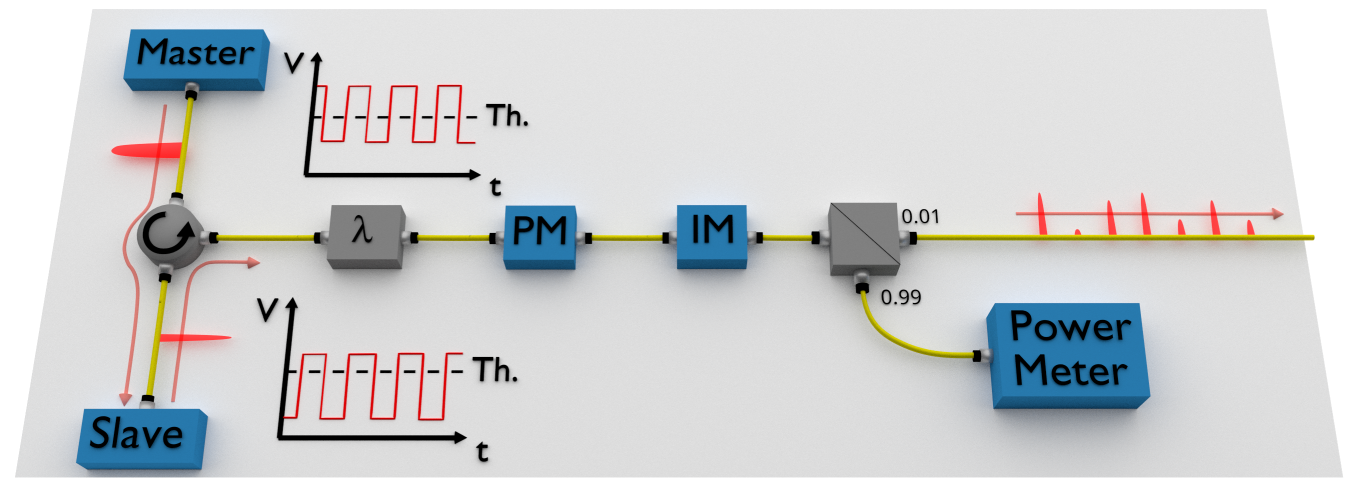}\\
  \caption{\textbf{Transmitting module.} Two replicas of the depicted fiber-based setup are used by Alice and Bob to transmit light pulses to Charlie. Master and slave lasers' driving signals are displayed alongside the equipment and are set differently for the two lasers. A power meter connected to the output beam splitter is used to attenuate the outgoing optical pulses to the correct level. The wavelength filter ($\lambda$) is for enhancing the indistinguishability of Alice's and Bob's photons.}
  \label{fig2}
\end{figure}

The preparation of the pulses in the experimental setup is effected through the transmitter depicted in Fig.~\ref{fig2}. Alice and Bob create low-jitter 32-ps light pulses at~1549.8~nm using the pulsed laser seeding technique~\cite{CLF+16}. The master laser is input to the slave via a circulator and the AC voltage is temporally offset between the two lasers to ensure injection occurs at the correct time.
The 1-GHz gain switching of both lasers ensures that all pulses are phase-randomised~\cite{JCS+11,YLD+14}. Alice and Bob's pulses are then passed through separate 30 GHz bandwidth filters to remove noise.
The polarization of the pulses is controlled using electric polarization controllers, which can create all of the required polarization states.
An attenuator provides the four photon fluxes ($s$, $u$, $v$, $w$) before they are sent to Charlie.

Charlie is composed of the interfering beam splitter (BS), two polarising BS (PBS) aligned along the $Z$ axis, one polarization rotator (R) and four InGaAs self-differencing avalanche photodiodes, run at room temperature, clocked at 1~GHz and featuring an average efficiency of 20.9\%. Overall, the setup shows similarities with a realisation of the original decoy-state MDI-QKD scheme~\cite{LCQ12} and the one realised in~\cite{CLF+16}. However, differently from previous implementations, the rotator R placed after one output port of Charlie's BS turns a $Z$-basis analyser into an $X$-basis analyser. This is the key element for enabling reconfigurable MDI/QKD as it allows the realization of a full QKD receiver, measuring the incoming pulses in the $Z$ and $X$ bases. On the other hand, it entails that all the coincidence counts from detectors $H$ and $V$ can be treated as in the original MDI-QKD scheme, whereas coincidence counts from detectors on different output ports of Charlie's BS cannot be used to distil key bits, as they belong to different bases. A coincidence count between $H$ and $V$ indicates projection onto the \textit{triplet} Bell state $|\psi^+\rangle=1/\sqrt{2}(|HV\rangle+|VH\rangle)=1/\sqrt{2}(|DD\rangle-|AA\rangle)$. In this case, Bob flips (does not flip) his bit to match Alice's bit if the rectilinear (diagonal) basis was used in the preparation step. The same argument applies to the $X$-basis branch, by replacing $H$ with $D$ and $V$ with $A$. Alice, Bob and Charlie share a common reference clock, allowing Alice and Bob to align their pulses, so to arrive coincidentally at Charlie, and allowing Charlie to align his detectors.

To effect the selection between MDI-QKD and QKD, Alice and Bob act on their IMs to send or stop the light directed to Charlie.
In particular, when Alice (Bob) prepares the vacuum state $w$, the amount of light travelling towards Charlie is so small that the situation is virtually identical to having the AC (BC) link disconnected and QKD enabled on the BC (AC) link. Any potential residual light in the vacuum state does not affect the security of the scheme, as it directly translates into an increase of the measured quantum bit error rate (QBER). Also, if there are multiple counts in Charlie's detectors during the QKD sessions, they can be treated using the squashing model for the passive BB84 protocol~\cite{squash1,squash2}. Finally, in some cases, neither Alice or Bob will prepare a vacuum state, whereas in other cases they both will. Such instances can be employed to enable an MDI-QKD communication back again. In other terms, the vacuum state acts as a switch between the different modalities of communication, QKD or MDI-QKD, thus enabling a reconfigurable MDI/QKD network (see details in~\cite{SM}).

\subsection*{Key rates}
\label{secsub:results}

\begin{figure}[t]
  \centering
  \includegraphics[width=\columnwidth]{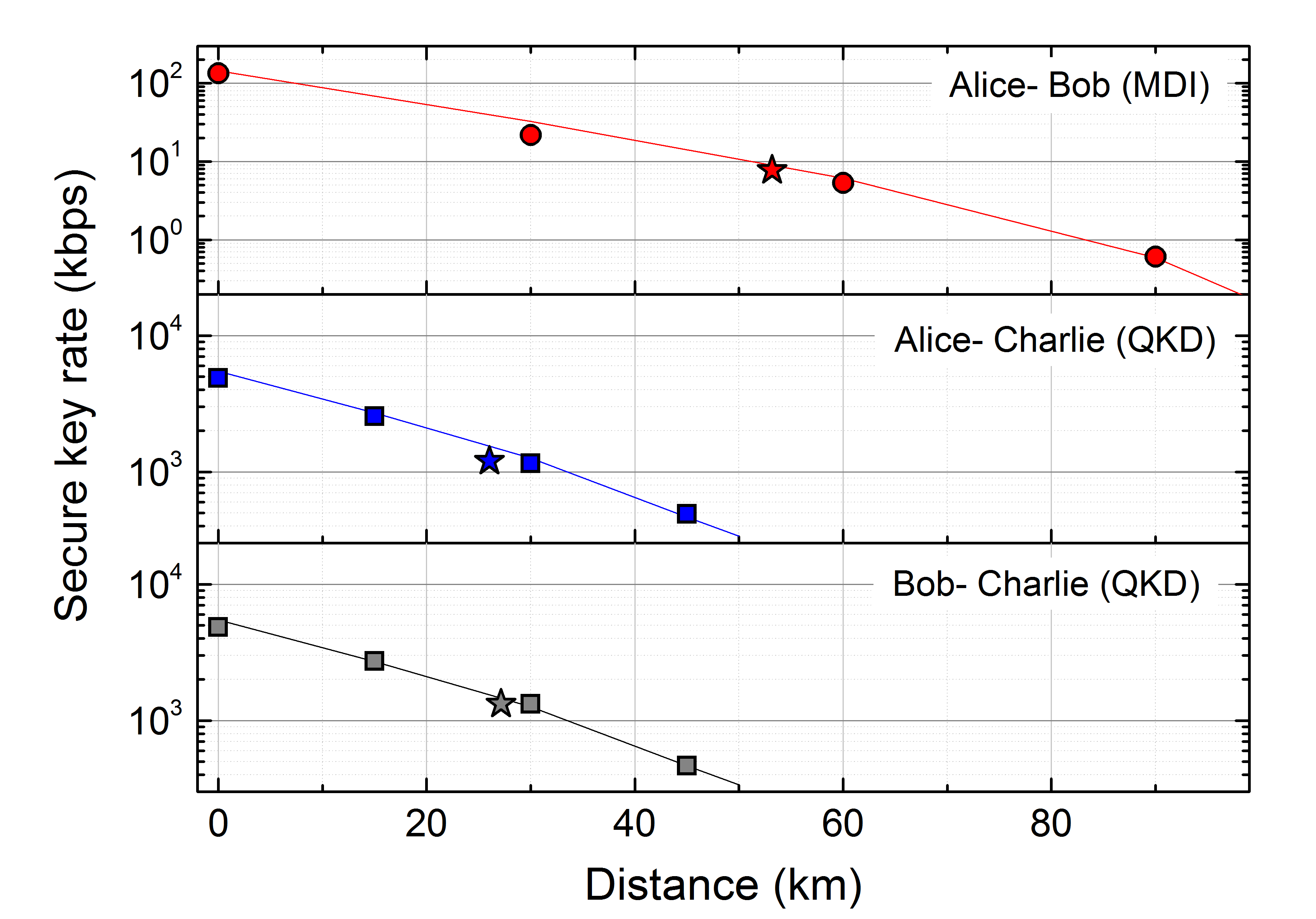}\\
  \caption{\textbf{Secure key rates versus distance}. MDI-QKD (top diagram) and QKD (centre and bottom diagrams) secure key rates for a $Z$ ($X$) basis probability equal to $80\%$ ($20\%$) and a security parameter $\epsilon_{\textrm{sec}} \lesssim 10^{-10}$ are shown as a function of distance. Because the reconfigurable scheme is symmetrically deployed, the distance between Alice and Bob (top) is approximately twice the distance between Alice and Charlie (centre) and Bob and Charlie (bottom). All the distances have been calculated assuming 0.2~dB/km attenuation on the channel, except for the data points indicated by stars, where a real fiber was used. The projected time to attain the shown key rates is 25 hours of which 3 minutes are spent on QKD and the rest on MDI-QKD, to balance their key rates.}
  \label{fig3}
\end{figure}

\noindent Using the described setup, we run QKD and MDI-QKD experiments, deriving key rates versus distance, as depicted in Fig.~\ref{fig3}. We performed two sets of experiments. In the first, we used variable optical attenuators to simulate a lossy channel with 0.2~dB/km, as in a typical optical fiber at 1550~nm. In the second, we used two 25~km reels of a standard optical fiber. The circles (squares) are for the attenuator-based MDI-QKD link (QKD links), whereas the stars represent the points obtained using a real fiber. Finally, the solid lines are theoretical simulations tailored to our experimental setup. Tables containing all the measured counts are reported in~\cite{SM}.

From Fig.~\ref{fig3}, it is apparent that the theory reproduces the experimental results well, both for attenuators and real fiber, with only a slightly lower experimental rate for the fiber due to a correspondingly higher QBER. This allowed us to use a simulation to optimise the system before performing real experiments. The key rates for MDI-QKD are between 606~bps for an equivalent distance of 90 km and 134~kbps at slightly more than 0~km. QKD is faster, providing key rates ranging from about 0.5~Mbps at 45~km to almost 5~Mbps at 0~km. This difference in the key rates led us to set the probability of an MDI-QKD run equal to 500 times that of a QKD run. In a network with three users, this would provide comparable key rates for all users, on average, over all distances. Before performing the experiment, the setup was optimised for MDI-QKD. Then, key rates were acquired for both MDI-QKD and QKD without additional calibrations.

Key rates in Fig.~\ref{fig3} were calculated using composable security proofs in the finite-size scenario~\cite{LCW+14,CXC+14} with a procedure similar to the one described in~\cite{CLF+16}. With the proviso that the key bits are extracted in the $Z$ basis, we drop the index $Z$ and write them as
\bea
\nonumber \textrm{R}^{\textrm{MDI}} &=& \underline{S}^{1,1} [1-h (\overline{e}^{1,1}_{\textrm{ph}})]-\textrm{leak}_{\textrm{EC}}^{\textrm{MDI}}-\Delta^{\textrm{MDI}}, \\
\textrm{R}^{\textrm{QKD}} &=& \underline{S}^{1} [1-h (\overline{e}^{1}_{\textrm{ph}})]-\textrm{leak}_{\textrm{EC}}^{\textrm{QKD}}-\Delta^{\textrm{QKD}}.
\label{eqnrates}
\eea
In Eq.~\eqref{eqnrates}, the labels ``MDI'' and ``QKD'' refer to MDI-QKD and QKD, respectively. The quantities $S$ and $e_{\textrm{ph}}$ indicate single-photon counts and single-photon phase-error rate, respectively, in the $Z$ basis and intensity class $s$, estimated by applying the decoy-state technique to the $X$-basis data sample and then extending to the $Z$ basis using standard statistical tools~\cite{SM}. The function $h$ is the binary entropy. The upper and lower bars are for upper and lower bounds and the superscripts `1' or `1,1' refer to one sender (QKD) or two senders (MDI-QKD) emitting single photons. The quantity $\textrm{leak}_{\textrm{EC}}$ is the amount of bits used to correct errors in the $Z$ basis, while the $\Delta$ terms take into account the finite-size effect.

\section*{Quantum digital signatures mediated by MDI-QKD}
\label{sec:qds}

\noindent Digital signatures play a vital role in software distribution, modern communication and financial transactions, where it is essential to detect forgery and tampering. Signatures are a cryptographic technique for validating the authenticity and integrity of messages, software, or digital documents.

\begin{figure}[t]
  \centering
  \includegraphics[width=\columnwidth]{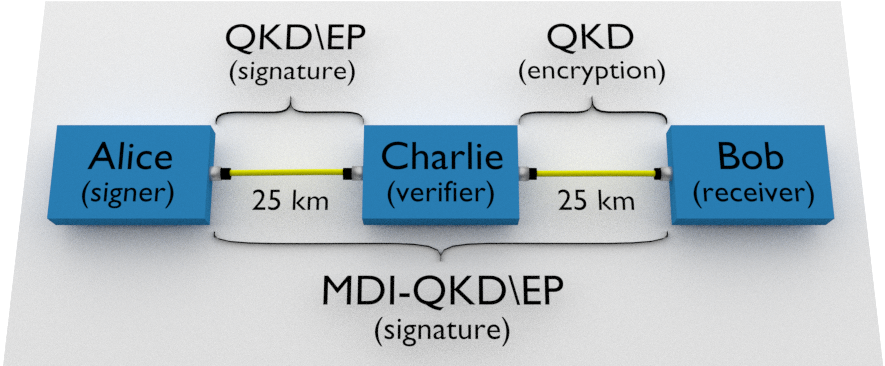}\\
  \caption{\textbf{MDI-QKD-mediated QDS}. The signature is sent by Alice to Bob using an MDI-QKD setup, over an optical fiber with a total length of 50~km. The protocol is denoted ``MDI-QKD$\backslash$EP'', where ``$\backslash$EP'' stands for ``\textit{without error correction and privacy amplification}''. QKD$\backslash$EP is used to send a signature from Alice to Charlie, whereas full QKD is used to distribute keys between Bob and Charlie to allow for the symmetrisation step of QDS (see~\cite{SM} for more details). The QKD links are implemented with two 25-km reels of single-mode fiber.}\label{fig4}
\end{figure}

In the simplest case, a digital signature scheme involves three parties, see Fig.~\ref{fig4}. One of them, Alice, signs a document and sends it to a receiver, Bob, who accepts it after checking that the signature is genuine. The same document can also be transferred by Bob to a third user, Charlie, for verification purposes. The users can also exchange their roles giving rise to similar schemes.
Differently from encryption, the goal of digital signatures is demonstrating the authenticity of a signed message to multiple recipients rather than keeping it secret. However, only the legitimate sender should be able to sign messages, to prevent forging of signatures by illegitimate users.

Digital signatures are currently implemented using public-key cryptography~\cite{DH76,RSA78} for which security depends on the computational power of an adversary. On the other hand, QDS  provide information-theoretic security based on the laws of quantum physics. While the seminal work on QDS described in~\cite{GC01} required the use of a quantum memory, recent developments have removed this restriction~\cite{ACJ06, DWA+14} and combined QDS with existing QKD and MDI-QKD schemes~\cite{AWK+16,PAW+16}.
Experiments have been performed for QDS up to distances of 90~km in optical fiber~\cite{CCD+12,CDD+14,CAF+16}.
However, this was achieved using a protocol secure only against individual attacks in the asymptotic scenario. More importantly, in all previous schemes, three or more physical channels were envisaged to implement the QDS protocol between three parties and none of them was assigned to MDI-QKD.

We implement QDS with only two optical channels using reconfigurable MDI/QKD. The adopted configuration is summarised in Fig.~\ref{fig4}. Two 25-km optical fibers connect Alice and Bob to Charlie whereas there is no direct fiber between Alice and Bob, who are linked only by the intermediate node using MDI-QKD. We choose this particular realization of QDS to demonstrate signature distribution based on MDI-QKD. However, given the key rates represented by stars in Fig.~\ref{fig3}, we could also have used the setup with the signing party at the ``Charlie" node, in which case MDI-QKD would be used to encrypt the symmetrising  exchange of signature bits between the two recipients at the ``Alice" and ``Bob" nodes.
Therefore our demonstration is not limited to the particular case represented in Fig.~\ref{fig4}. Let us also remark that, to strengthen our demonstration, we used real fiber to perform the QDS experiment.

To demonstrate QDS mediated by MDI-QKD, we employ a protocol that combines those presented in~\cite{AWK+16} and \cite{PAW+16}. It adopts MDI-QKD without error correction and privacy amplification (denoted by ``MDI-QKD$\backslash$EP'') for the signature between Alice and Bob, QKD$\backslash$EP for the signature between Alice and Charlie, and full QKD between Bob and Charlie to enable the symmetrisation step of the protocol (see~\cite{SM} for details). All the results in the $X$ basis are publicly revealed, whereas only a small portion of the $Z$ basis results are disclosed, to estimate the QBER in this basis. The remaining undisclosed bits are used for quantum digital signatures.

\subsection*{Signature rates}
\label{secsub:resultsQDS}

\noindent Here we extract the specific parameters of a QDS protocol, i.e., the size of signatures and the security parameters against forging and non-repudiation~\cite{AWK+16,PAW+16}.
For that, we introduce a protocol with a different finite-size treatment than previous protocols, to increase the QDS rate.

In the finite-size scenario, the users acquire data for a finite amount of time, until the data block is large enough to guarantee small statistical fluctuations in the parameters estimated from the data set. Then they proceed and acquire the next block of data. The current approach for QDS is to distil a single signature from every block of data. Therefore, to increase the QDS rate, it is optimal to keep the data block as small as possible, so as to have more blocks in a given time interval. This, however, makes statistical fluctuations larger, thus worsening the estimation of the quantum-related parameters and the QDS rate.

In our protocol, we still perform a single decoy-state parameter estimation per each block of data. However, we consider the extraction of multiple signatures from the same data block. This allows us to acquire a large data set, minimising the statistical fluctuations, and at the same time distil as many signatures as possible from each acquired block. We estimate that this improves the QDS rate by about 10 times over the standard approach, depending on the experimental conditions.

We start from analysing the MDI-QKD-mediated QDS rate on the link connecting Alice and Bob. Then we will apply an analogous procedure in order to estimate the QDS rate on the QKD link between Alice and Charlie. The QDS protocol also includes a symmetrization step between Bob and Charlie performed on a secure channel~\cite{AWK+16,PAW+16,SM}. This can be enabled by running QKD on the remaining link between Bob and Charlie. The specific key rate for this scheme would be the one showed in the bottom diagram of Fig.~\ref{fig3}.

As a first QDS-specific quantity, we evaluate the minimum rate $p_E^{\textrm{MDI}}$ at which Eve can introduce errors on the MDI-QKD link. This is given by
\beq
h \left( p_E^{\textrm{MDI}} \right) =\underline{S}^{1,1}_{\textrm{sig}} / C^{s,s}_{\textrm{sig}} [1-h (\overline{e}^{1,1}_{\textrm{ph{ ,sig}}})],
\label{pE}
\eeq
which is derived from Eq.~\eqref{eqnrates}, omitting the error correction and, for simplicity, the finite-size terms~\cite{PAW+16}. In Eq.~\eqref{pE}, the subscript `sig' indicates that the quantities refer to the block from which signatures are extracted. In the QDS protocol, Alice randomly selects $C^{s,s}_{\textrm{sig}} = 2.5 \times 10^6$ bits from the $Z$-basis block, to form one of the signature blocks. Because this size is smaller than the Z-basis dataset, she will be able to extract multiple signature blocks from it, all with size $C^{s,s}_{\textrm{sig}}$.
She then applies decoy-state estimation to find, for the signature block, the lower bound for Charlie's counts due to single-photon pulses and the corresponding upper bound for the phase-error rate. In our experiment the two bounds are $\underline{S}^{1,1}_{\textrm{sig}}= 666,345$ bits and $\overline{e}^{1,1}_{\textrm{ph,sig}}= 0.053$, respectively, leading to $p_E^{\textrm{MDI}}=0.0286$.

The next step is to determine an upper bound for the QBER in the signature blocks, $\overline{E}^{s,s}_{\textrm{sig}}$. For that, the QBER was directly measured on a sample set of $C^{s,s}_{\textrm{test}}=1,714,426$ bits and found to be ${ E^{s,s}_{\textrm{test}}}=0.5\%$ (see also Table~4 in~\cite{SM}). The measured block can be thought of as a random sample drawn from the overall $Z$-basis population. Therefore the measured QBER is representative of the QBER in the non-measured fraction of the population. From this fraction, the users select several blocks of size $C^{s,s}_{\textrm{sig}}$ to form the signatures. The QBER in each signature block is then estimated, by applying Serfling's inequality~\cite{Ser74}, to be
\beq
\overline{E}^{s,s}_{\textrm{sig}} = E^{s,s}_{\textrm{test}} + \sqrt{\frac{ (C_{\textrm{sig}}^{s,s} + 1) (C_{\textrm{sig}}^{s,s} + C_{\textrm{test}}^{s,s})}{2 ~C_{\textrm{test}}^{s,s}   (C_{\textrm{sig}}^{s,s})^2   } \ln (1/\ep_H)}.
\label{hoeff}
\eeq
The estimation in Eq.~\eqref{hoeff} provides $\overline{E}^{s,s}_{\textrm{sig}}=0.0085$ when we set $\ep_H=2\times10^{-11}$. After calculating suitable authentication and verification parameters, $s_{ab} = 0.0152$ and $s_{vb} = 0.0219$, we obtain the length of a signature $L_{\textrm{sig}}^{\textrm{MDI}}$ by inverting the relation
\beq\label{Prep}
P^{\textrm{MDI}}_{\textrm{rep}} \leq  \exp \left[ -(s_{vb} - s_{ab})^2 L_{\textrm{sig}}^{\textrm{MDI}} /4 \right] \leq 0.5 \times10^{-10},
\eeq
which sets the repudiation probability $P^{\textrm{MDI}}_{\textrm{rep}}$ to less than $0.5 \times10^{-10}$~\cite{AWK+16,PAW+16}. The resulting value for $L_{\textrm{sig}}^{\textrm{MDI}}$ is $2.11\times 10^6$, which is smaller than the set value of $C^{s,s}_{\textrm{sig}}$, showing that Eq.~\eqref{Prep} holds in our experiment when we take $C^{s,s}_{\textrm{sig}}$ as the signature length. The overall failure probability at the end of the QDS distillation is less than $10^{-10}$.

A signature of size $C^{s,s}_{\textrm{sig}}$ can be generated with our system in $45$ seconds on average. This is a remarkable speed for MDI-QKD-mediated QDS, if the increased security level entailed by measurement-device-independence is taken into account. The average time results from the ratio of the total acquisition time in an experiment with an 80:20 bias between the $Z$ and the $X$ bases divided by the $1,974$ different signatures generated from the acquired data block. The reported average time includes the QKD operations on the other two links.
The analysis of MDI-QKD-based signatures is completed by calculating the probabilities of honest abort, $P_{\textrm{hab}}$, and forging, $P_{\textrm{for}}$, which are confirmed to be much smaller than the set threshold $10^{-10}$ with our experimental parameters.

As an additional step in the QDS scheme, we now evaluate the QDS rate on the 25-km QKD link between Alice and Charlie (see Fig.~\ref{fig4}).
We repeat similar calculations as for the MDI-QKD link. We set the size of the signature block to $C^{s}_{\textrm{sig}} = 150,000$ bits, randomly selected in the $Z$-basis data block acquired by operating QKD on the AC-link.
In the signature block, the lower bound for Charlie's counts due to single-photon pulses amounts to $\underline{S}^{1}_{\textrm{sig}} = 86,563$ and the upper bound for the phase-error rate is $\overline{e}^{1}_{\textrm{ph,sig}}=0.0237$, leading to $p_E^{\textrm{QKD}}=0.105$.

The actual QBER in the $Z$ basis was measured on a sample of $46,979,354$ bits and amounts to $E^s_{\textrm{test}}=0.0017$ (see also Table~3 in~\cite{SM}). From it, using an equation similar to Eq.~\eqref{hoeff}, we obtain an upper bound $\overline{E}^s_{\textrm{sig}}=0.0108$, which is less than $p_E^{\textrm{QKD}}$, thus providing a positive QDS rate. To determine the rate, we calculate $s_{ac} = 0.0421$ and $s_{vc} = 0.0734$ and obtain a signature length $L_{\textrm{sig}}^{\textrm{QKD}}=103,336$ by inverting an equation similar to Eq.~\eqref{Prep}, but with $p_{\textrm{rep}}^{\textrm{QKD}}$ replacing $p_{\textrm{rep}}^{\textrm{MDI}}$. The total repudiation probability is then given by the sum $p_{\textrm{rep}}^{\textrm{QKD}}+p_{\textrm{rep}}^{\textrm{MDI}}$.

The total time the system would spend acquiring QKD data on the AC link is about 36 seconds. The resulting data block would be enough to distil signatures for 2,506 1-bit messages, thus providing an average time for the QKD-only operations on the AC link equal to 72~ms for each signed bit.
A similar value of 74~ms could have been obtained on the QKD link between Bob and Charlie if we had used it to distil signatures rather than for encryption. Although the reported values can be further improved by optimising the initial parameters set by Alice and Bob, they are already in line with state-of-the-art QDS, especially when considering the fact that the present scheme offers security in the finite-size scenario against the most general attack allowed by the laws of physics.

\section*{Conclusion}
\label{sec:conclusion}

\noindent Fig.~\ref{fig1} shows a small quantum network, where Charlie can either act as an ``end user'' or as a relay between the other two users. In the former case, he is a legitimate trusted party; this could be the case e.g. if the setup is used for standard QKD between Alice and Charlie. In the latter case, Charlie might not be trusted, but the MDI-QKD protocol guarantees that he cannot compromise the security of the overall network. This is particularly important in a QDS distribution scheme, where all parties need to cooperate to distil valid signatures, but where none of the participants are fully trusted.

From a security point of view, the key distributed between Alice and Bob via the MDI modality is the most secure and is also guaranteed against any attack directed at Charlie's detectors. This guarantee does not apply to the QKD modality. However, the QKD modality is much faster and can serve different, rate-hungry, applications.
Interestingly, the considered reconfigurable network only uses two optical links to distribute quantum keys and signatures between three users.
This scales favourably. If the relay is the central node of a star network, quantum strings can be distributed among $N$ parties using only $N-1$ optical links, as opposed to directly connecting all the parties with $N(N-1)/2$ links. While this is common to all star networks, it should be stressed that in this case the central node is totally untrusted, a positive feature due to the MDI-QKD protocol.

The above-described scheme allows to switch between QKD and MDI-QKD in real time, using exactly the same equipment for QKD as in MDI-QKD, with the exception of a single rotator added to Charlie's module. We have experimentally demonstrated its practicality by obtaining QKD and MDI-QKD key rates as well as the first QDS rates mediated by MDI-QKD.
In the QDS experiment, about 45 seconds and 72~ms were required, on average, to acquire enough signature material to sign 1 bit from the MDI-QKD protocol over a 50~km optical fiber and from the QKD protocol over a 25~km optical fibre, respectively.
To increase the efficiency of the distillation, we have introduced an original treatment of the finite-size effect, detailed in~\cite{SM}, where multiple signatures were distilled from the same block of data, thus alleviating the detrimental effect of statistical fluctuations.

\textit{Acknowledgements} - We acknowledge Ryan Amiri and Petros Wallden for useful discussions. G.L.R. gratefully acknowledges financial support from the EPSRC CDT in Integrated Photonic and Electronic Systems and Toshiba Research Europe Ltd. M.C. acknowledges support from the Galician Regional Government (Grant No. EM2014/033, and consolidation of Research Units: AtlantTIC), the Spanish Ministry of Economy and Competitiveness (MINECO), the Fondo Europeo de Desarrollo Regional (FEDER) through Grant No. TEC2014-54898-R, and the European Commission (Project QCALL). I.V.P. and E.A. acknowledge financial support by EPSRC grant EP/M013472/1.

\end{document}